# ELECTRONIC-GOVERNMENT IN SAUDI ARABIA : A POSITIVE REVOLUTION IN THE PENINSULA

**Omar S. Al-Mushayt[1*], Kashiful Haq[2] and Yusuf Perwej[3]**
Computer Science & Information System
Jazan University, Jazan, KSA
E-mails : kashiful@rediffmail.com[2], yusufperwej@gmail.com[3]

## ABSTRACT

The informatization practice of countries all over the world has shown that the level of a government's informatization is one main factor that can affect its international competitive power. At present, e-government construction is regarded as one of the most important tasks for the national economy and society upliftment and informatization in Saudi Arabia. Unlike the traditional governments, an e-government takes on a new look with its framework and operation mode more suitable for the contemporary era. In fact, it is a basic national strategy to promote Saudi Arabia's informatization by means of e-government construction. This talk firstly introduces the basic concepts and relevant viewpoints of e-government, then reviews the development process of e-government in Saudi Arabia, and describes the current states, development strategies of e-government in Saudi Arabia. And also review e-government maturity models and synthesize them e-government maturity models are investigated, in which the authors have proposed the Delloite's six-stage model, Layne and Lee four-stage model and Accenture five-stage model. So, the main e-government maturity stages are: online presence, interaction, transaction, transformation and digital democracy. After that, according to many references, the main technologies which are used in each stage are summarized.

## INTRODUCTION

E-Government has been employed by developed as well as developing countries to be an enabler toward accelerating processes, delivering a higher level of service to citizens and businesses, increasing transparency and accountability while lowering costs. There are a number of definitions for e-government in the literature. A popular definition which is David McClure's view [McClure (2000)] says: "Electronic government refers to government's use of technology, particularly web-based Internet applications to enhance the access to and delivery of government information and services to citizens, business partners, employees, other agencies and entities". E-government is a way of making government smarter and smaller, improving delivery of services and giving citizens new tools to interact with government. Therefore citizens can expect better, cheaper, faster and more accessible services [Turban et al. (2002)] E-government includes:

1. Providing greater access to government information.
2. Promoting civic engagement by enabling the public to interact with government officials.
3. Making government more accountable by making its operations more transparent and thus reducing the opportunities for corruption and,
4. Providing development opportunities, especially benefiting rural and traditionally underserved communities.

* Corresponding author, E-mail : oalmushayt@yahoo.com

E-Government is a special type of electronic business with particular objectives and characteristics. It utilizes the internet and web-based technologies to provide government services online to citizens, businesses, and other government agencies so as to bring about economic benefits [Andersen and Henriksen (2008)].

E- Government can provide substantial benefits for citizens, businesses, and governments around the world [Jaeger and Thompson (2007)]. E-Gov has been promoted as a key to "radically shrinking communications and information costs, maximizing speed, broadening reach, and eradicating distance".

E-Government has emerged worldwide as a trend to offer electronically administrative service packages that meet the needs of citizens' life events and business transactions, with a promise to enhance service accessibility and alleviate service delivery delays and costs. [Dimitris Gouscos, Manolis Kalikakis and Maria Legal (2007)].

**OBJECTIVE OF E-GOVERNMENT**

The objective of e-Government is not just to computerize government offices, it is to gradually transform the way the government operates. E-government refers to the use by government agencies of information technologies that have the ability to transform relations with citizen, businesses, and other arms government. But the process will take time and significant amount of re-engineering of processes. Hence, e-Government is not just another way of doing existing activities; it is a transformation on a scale that will fundamentally alter the way public services are delivered. It does not have a time-line; rather it is evolutionary. The relationship is no longer just a one-way, Government versus them citizens proposition; rather it is about building a partnership between governments, and their citizens. Through transformation/ re-engineering of processes, Government should achieve the following
1. Better delivery of government services to citizens,
2. Improved interactions with businesses and industry,
3. Empowering citizens through access to information or more efficient government management,
4. Increased transparency and greater convenience,
5. Reduce corruptions, and costs.

**SOME TANGIBLE BENEFITS OF E-GOVERNMENT**
1. **Increases transparency:** E-Government allows increased transparency of government activities and makes the government more accountable to citizens.
2. **Reduces scope for corruption:** Reduced scope for corruption is another important impact of e-Government. Combating corruption is a top priority and e-government can provide an effective tool for that purpose.
3. **Helps increase investor confidence:** Improvements in the transparency of government also raises investor confidence, which in turn contributes to increased foreign direct investment in the long run.
4. **More efficient governance:** E-Government facilitates in making the government's internal processes more efficient, thus saving time and resources in the long run.
5. **More efficient services to citizens:** E-Government enables the government to respond more efficiently and quickly to citizen demands and requests.
6. **Helps boost the private sector:** E-government helps provide boost to the private sector, particularly SMEs, by reducing the time and expense required for businesses to interact with the government. Furthermore, through simplification of government processes and services such as online procurement, the government can reduce barrier to entry for new businesses and also increase competition.
7. **Allows for decentralization of governance:** E-Government makes decentralization of government easier since data stored in digital format can be updated and accessed from virtually any office within a networked environment.
8. **Allows greater scope for integration:** Digital storage of data and software applications allow greater scope of integration of activities of different government offices as data can be shared easily and efficiently.
9. **Allows learning from the past:** Since e-Government allows data to be stored and from past projects can be easily used for new retrieved easily, experiences and statistics similar projects.

10. **Stimulates the local ICT industry:** e-Government projects also provide valuable experience to the local ICT industry for becoming competitive in an international market.
11. **Makes ICT relevant to the masses:** E-government makes ICT relevant to the masses as its benefits can gradually be shared by all from every corner of the country.

## COMPONENTS OF THE E-GOVERNMENT FRAMEWORK

Any successful e-government initiative is concerned with providing improved government e-services to the user, be it an individual, a business or a government agency. Moreover, it also aims to increase efficiency and effectiveness [Yasser (2006)] government's administration. In order to be able to do so, an e-government initiative has to address several issues and has to put into place several components of an e-government framework.

### Components of E-Government Framework for Saudi Arabia

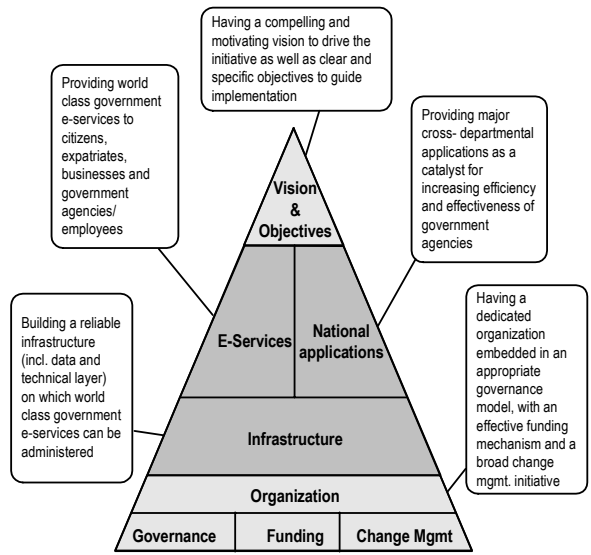

## VISION STATEMENT FOR SAUDI ARABIA'S E-GOVERNMENT INITIATIVE

The vision for Saudi Arabia's e-government initiative is user-centric and focuses on a number of aspects which all revolve around the central notion already mentioned, i.e., providing better government services to the user. As mentioned before, users are understood here as individuals (citizens and expatriates), businesses and government agencies. The user-centric vision for Saudi Arabia's e-government initiative is summarized by the following vision statement [Yasser (2006), Saudi Computer Society (2004)].

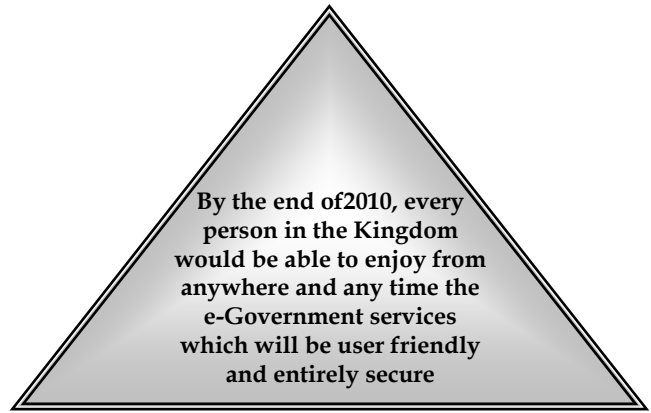

**By the end of 2010, every person in the Kingdom would be able to enjoy from anywhere and any time the e-Government services which will be user friendly and entirely secure**

## STRATEGIC OBJECTIVES OF SAUDI E-GOVERNMENT INITIATIVE

The vision for Saudi Arabia's e-government initiative was detailed by determining ten objectives to be achieved within the end of the year 2010 [Yasser (2006)]. Those objectives are divided as follows:

**Provide better services by the end of 2010**

1. Provide the top priority services (150) at world class level of quality electronically.
2. Deliver services in a seamless and user friendly way and at highest standards of security.
3. Make services available to everyone in the Kingdom and allow 24/7 access from cities as well as countryside and even outside the country.
4. Realise 75% adoption rate with respect to the number of users.
5. Ensure 80% user satisfaction rating for all services provided electronically.

**Increase Internal Efficiency and Effectiveness**

6. Deliver all possible official intra-governmental communication in a paperless way.
7. Ensure accessibility of all information needed across government agencies and storage of information with as little redundancy as possible.
8. Purchase all goods and services above a reasonable value threshold through e-procurement.

**Contribute to Country's Prosperity**

9. Contribute to establishment of information society in the kingdom through spreading information, knowledge and use of e-services.
10. Help improve use of country's assets and resources by increasing society's productivity in private, business and public sector.

## RESEARCH METHODOLOGY

Meta-Synthesis approach is used to produce interpretive translation, ground narratives or theories by integrating and comparing the funding or metaphors of different qualitative studies [Beck (2002)]. This method is used to integrate multiple studies in order to produce comprehensive and interpretive finding by comparing, interpreting, translating and synthesizing different research frameworks. This method has been used in social sciences and medical area. This research adopts Noblit and Hare (1988) seven-step approach. These seven steps are: getting started, searching and selecting of relevant studies, reading the studies, determining how the studies are related, translating the studies into one another, synthesizing translations and expressing the synthesis.

## PROCEDURE OF RESEARCH

As mentioned before, in this research, meta-synthesis approach is used that followed the seven-step Meta–ethnography of Nobilt and Hare (1988).

1. **Getting started:** The aim of this research is study on e-government maturity stages.
2. **Select relevant studies:** the current literature related to e-government maturity models searched and one model was identified.
3. **Reading the studies:** one maturity models was reviewed and details of each stage was investigated.
4. **Determining how the studies are related:** In this steps, relationship between different studies are shown. Study of these models showed that developing trends of these models are similar, although they are based on various perspectives.
5. **Translating the studies into on another:** In this step, this research deals with compares different models and finds their corresponding relationships. The stages of models can be translated to each other. For example, the first stage of models, provide public information on line.
6. **Synthesizing translations:** this step, show the complexity, time-taken, and level of integration increase with each succeeding stage.
7. **Presenting the finding:** in this stage, result of research and finding organized into text and diagram.

**OVERVIEW OF E-GOVERNMENT MATURITY MODELS**

A citizen-oriented strategy cannot be achieved by only putting processes on the Internet and launching the websites, although this is the first and essential step. E-government is far more than just websites. E-government is about business transformation and may involve a government reform . To achieve a fully functional e-government which can perform all the interactions and transactions online, the system should evolve gradually while all required facilities are being prepared. This results in a step-wise completion of the system. The stages of development start from publishing the information online to a website with full transactions between different departments of the government. The stage-model outlines the available services and structural transformations of governments as they progress towards an electronically-enabled government. This progress may imply fundamental changes in the form of government. In this section, three e-government maturity models are described which some of them developed by individual researchers and other ones by institutions.

### Delloite's six-stage model

Delloite group proposed a maturity model for e-government which contains six-stage as follows:
1. Information publishing: Each government department sets up its website to provide information about itself.
2. "Official" two-way transaction: Users can transact information to individual departments with secure websites.
3. Multipurpose portal: A portal allows customer to use a single point of entry to send and receive information.
4. Portal personalization: Users can customize the portal with their desired features.
5. Clustering of common services: Real transformation of government structure takes shape. All services will be clustered along common lines by government.
6. Full interaction and enterprise transformation: The structure of government is changed and technology is integrated across the new structure.

### Layne and Lee four-stage model

Layne and Lee (2001) proposed a four-stage model for e-government as follows:
1. Cataloguing: Includes on-line presentation and ability of downloadable forms.
2. Transaction: Putting live database links to on-line Interfaces.
3. Vertical integration: Online integration between different levels of government.
4. Horizontal integration: On line integration between different levels and different functions of government

**Accenture five- stage model**

Accenture (2007) in its report considered five stages for e-government development as follows:

1. Online presence: Includes publishing government information online, making available a few services like downloading forms. This stage needs lowest amount of technical ability.
2. Basic capability: Includes creating a central plan, developing a legislative framework, addressing the security and certification problem, broadening the online presence, and implementing some easy transaction capabilities. Customers can submit their personal information to individual agencies and digital signature is introduced. This stage needs higher level of technical abilities.
3. Service availability: A basic portal website should be available. This is the gateway to the e-government system. Some efforts made to integrate services available through different agencies. Some cross – agencies or horizontal cooperation started, and citizen focus is presented.
4. Mature delivery: Clear ownership, responsibility and authority, intra-agency relationships and collaboration across different levels of government should be implemented. Stronger services with the ability of adding more value are considered. The issue of customer service is emerging to the system.
5. Service transformation: The main vision is to improve customer service by removing any problems the user are facing. The e-government can deliver all the services a government usually provides in online manner.

**THE CHALLENGES FACED**

In order to achieve a complete solution, one that increases citizen participation and engagement through effective use of ICT technology, governments are faced with numerous challenges:

1. For the population at large they must provide an opportunity for equitable and affordable access to high speed Internet services in order to address the socio-economic status issues of ICT today, commonly referred to as the digital and technology divide.
2. In order to accelerate the acceptance and usage of e-Government services, both today and future services, they must find a provisioning method that is sustainable, scalable and practical in terms of being economically viable.
3. They must create a business environment that will encourage private industry investment, collaboration and alliances.

4. They must create a marketing environment based upon a "Community-of-Interest" approach that will generate revenue for re-investment so that services can expand and evolve and the community will benefit in terms of economic, social and cultural development.

**Table I. Contributions of e-government models to identify main stages**

| Stage → <br> Model ↓ | Online presence | Interaction | Transaction | Vertical integration | Horizontal integration | Full integrated | Digital democracy |
|---|---|---|---|---|---|---|---|
| Delloite | √ | √ | √ | | | √ | √ |
| Layne & Lee | √ | | √ | √ | √ | √ | |
| Accenture | √ | √ | √ | | | √ | |

5. They need to ensure that the solution is "performance-based" addressing the key categories of financial, economic development, reduction of redundancy, fostering democratic principle, improving service to citizens and other constituencies.
6. Finally, they must find a solution that offers multiple agencies at all levels of government an opportunity for convergence, consolidation, and collaboration.

## DISCUSSION

By translating different stages to each other, table 1 is concluded. According to table 1 main stages in e-government maturity can be summarized as

1. Online presence: In this step, government starts toward e-government and publish useful information online.
2. Interaction: Government go further and citizen can interact with government by downloading forms, e-mailing to officials.
3. Transaction: In this step, typical services such as tax filling and payment, driver's license renewal are available
4. Fully integrated and transformed e-government: In this stage, delivery of government services is redefined by providing a single point of contact to constituents.
5. Digital democracy: Some services such as online voting, online public forums and online opinion surveys are available.

## BENEFITS OF CENTRALIZED DIGITAL COMMUNITY CENTERS OF GOVERNMENT (ALL LEVELS)

1. Provide full accessibility to Government Programs, reaching out to segments that heretofore have been unreachable.
2. Affordable public access to the Internet for all citizens offering single window access to e-Gov.
3. Benefits of Marketing to drive adoption, penetration and utilization – broaden reach for government programs.
4. Offer opportunity for government interagency collaboration, consolidation and convergence.
5. Encourages the development of innovative content and applications.
6. Minimizes deployment cost.
7. Responds to community needs.
8. Builds community capacity and creates a broader base of Internet users – market for future services and premium services and the benefits of economies of scale plus economies of scope.
9. Capabilities to ensure communities/citizens are able to stay current due to commercial nature of program.
10. Encourages development and diffusion of enabling technology, infrastructure and backbone.
11. Drives job creation, knowledge enhancement and workforce education.
12. Stimulates SOHO and SME growth at the community level and beyond.
13. Provides customer satisfaction.
14. Moves government from Website development to "interactive relationships" with the citizens.
15. Overcomes the limitations to Internet adoption – the socio-economic status (Digital Divide).
16. Position the government as an integrated enterprise.
17. Efficient and effective information delivery.
18. Broad based feedback mechanisms for government.
19. A positive business case for service delivery which is sustainable, scalable and cost effective and drives economic growth.

## TECHNOLOGIES USED IN E-GOVERNMENT

The main technology which is used in each stage of e-Government is summarized in figure1. According to Figure 1, there are different technologies for each stage. In online presence stage, just basic web technologies are used. But in the final stages, complex technologies is used such as secure communication network public key infrastructure

## CONCLUSION

In the final, the authors conclude that this research paper helps us to understand and also enhance further the tangible benefits of e-Governance, and also clarify the initiatives taken by the e-government of Saudi Arabia to improve the current status and future vision of improving its services given to its citizens. Also we conclude that by the end of 2010, the services will be better, and it will increase the internal efficiency & effectiveness, and contribute to its country's prosperity. As well as the paper proposed a e-government maturity model with five stages as online presence, interaction, transaction, transformation and digital democracy. Qualitative Meta synthesis approach is used to compare different e-government maturity models and synthesis them. The main technologies which are used for each stages are summarized by using meta synthesis approach too.

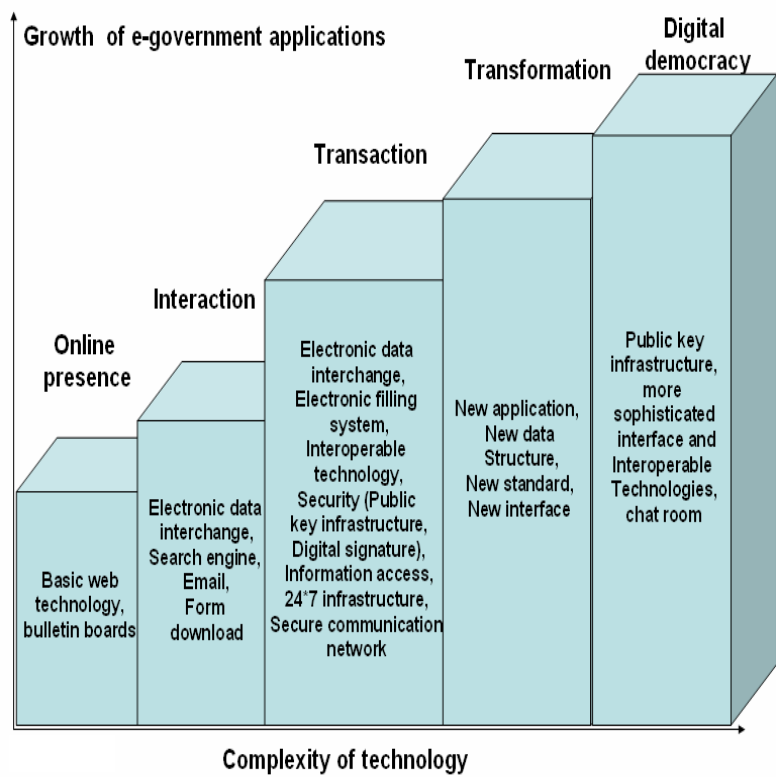

**Figure 1. E-Government maturity model & useful technologies in each stage.**